\documentclass[aps,prb,showpacs,twocolumn,groupedaddress]{revtex4}
\usepackage{graphicx}
\usepackage{epsfig}
\usepackage{amsmath}
\usepackage{amssymb}

\newcommand{\one}{\mbox{$1 \hspace{-1.0mm}  {\bf l}$}}

\newcommand{\Eqref}[1]{Eq.~(\ref{#1})}
\def\ket#1{\left| #1\right>}

\def\ketbra #1#2{\vert #1\rangle \! \langle #2\vert}
\newcommand{\proj}[1]{\ketbra{#1}{#1}}

\newcommand{\up}{\hspace{-3 pt}\uparrow}
\newcommand{\dn}{\hspace{-3pt}\downarrow}

\begin{document}

\title{Entanglement Generation via a Completely Mixed Nuclear Spin Bath}
\author{H. Christ}
\author{J. I. Cirac}
\author{G. Giedke}
\affiliation{ Max--Planck--Institut f\"{u}r Quantenoptik,
Hans-Kopfermann--Str. 1, D--85748 Garching, Germany }
\date{\today}

\begin{abstract}
  We show that qubits coupled sequentially to a mesoscopic static completely
  mixed spin bath via the Heisenberg interaction can become highly entangled.
  Straightforward protocols for the generation of multipartite entangled
  (Greenberger-Horne-Zeilinger-)states are presented. We show the feasibility
  of an experimental realization in a quantum dot by the hyperfine interaction
  of an electron with the nuclear spins.
\end{abstract}

 \maketitle
%--------------------
\section{Introduction}\label{sec:intro}
%---------------

The quest to realize quantum information processing (QIP) has
motivated an impressive race to implement high precision
preparation and manipulation of isolated two-level quantum systems
(qubits) in a wide variety of physical settings \cite{Zoller2005}.
A hallmark achievement for each such approach is the generation of
quantum entanglement through controlled interaction between two or
more qubits. Since switchable direct interactions between qubits
often entail additional decoherence mechanisms, many QIP proposals
rely on interactions \emph{mediated} by an additional quantum
system. %\cite{CiZo95,Kan98,IAB+99}
As a rule this mediator (just as the qubits themselves) needs to
be prepared in a pure state to achieve high-fidelity quantum
operations and it may look futile to use a high-entropy mesoscopic
spin bath for this task.  In contrast to these expectations, we
show here that high-fidelity entanglement generation can be
realized even if the qubits can only interact with an arbitrarily
mixed spin bath, provided that this interaction can be switched on
and off, single-qubit unitaries are available, and the bath has
slow internal dynamics. This is motivated by and will be
illustrated through the example of electron-spin qubits in quantum
dots (QDs) \cite{LoDi98}, where the ensemble of lattice nuclear
spins represents a strongly coupled but slowly evolving spin bath.

Nuclear spins in quantum dots have received much theoretical
\cite{KLG02,ErNa04,DeHu06,WiDa06,SKL03,YLS05} and experimental
\cite{PJT+05,KFE+05} attention in the QIP context as the main
source of electron-spin decoherence through the strong hyperfine
coupling. It has also been noted that their slow internal dynamics
and long (expected) decoherence time \cite{PLSS77} make the
ensemble of nuclear spins useful as a quantum memory \cite{TIL03}
or for quantum computation \cite{TGC+04}. These applications,
however, require careful and yet unachieved preparation of the
nuclear system. What we show here is, that the unprepared, highly
or even maximally mixed (nuclear) system is able to mediate
coherent interaction between electrons and thereby allows the
generation of highly entangled states of many (electron spin)
qubits without any electron-electron interaction.

We consider a QD in the single-electron regime \cite{HKP+07} and assume the
availability of single-electron state preparation and measurement as
well as the controlled shuttling of prepared electrons into and out
of the QD, all of which have been demonstrated experimentally
\cite{HaAw08}. Additionally required is control of the detuning (e.g.,
by a magnetic or electric field), which switches the hyperfine (HF)
interaction between resonant and off-resonant regimes. We first show
how sequential interaction of three electrons with the nuclear bath
can generate a maximally entangled pair of electron spins. More
generally, the class of states that can be generated via the spin
bath is characterized in terms of matrix product states. Finally we
show that imperfect electron spin operations, inhomogeneous
couplings between electron and nuclei and modifications to the ideal
static spin bath still allow for the scheme to be realized. In
situations where spin-orbit coupling is large, our scheme can be an
interesting alternative to the standard exchange based setups,
because it does not involve occupation of any higher orbital
levels~\cite{HuDa06,CoLo07}.
%---------------
\section{Entanglement Generation}\label{sec:protocols}
%---------------

We consider each electron coupled via the uniform Heisenberg
interaction to the bath of $N$ nuclear spins and to an external
magnetic field $B_z$ ($\hbar=1$)
\begin{equation}
H=\frac{A}{2N}\left(I^+S^- + S^+I^-\right) + \frac{A}{N}
I^zS^z+g^\ast \mu_B B_z S^z.\label{Hamiltonian}
\end{equation}
${\mathbf{S}}$ is the spin operator for the electron and $I^\mu =
\sum_i I^\mu_i$ are the three components of the collective nuclear
spin operators ($\mu=\pm,z$ and $[I^+,I^z] = -I^+ ,\,\, [I^+,I^-]
= 2 I^z$). $g^\ast$ is the electron $g$-factor and $\mu_B$ the
Bohr magneton. We consider spin-1/2 nuclei, neglect bath dynamics,
the bath spins' Zeeman energies, and inhomogeneities in the
Heisenberg couplings for now, and discuss the validity of these
approximations towards the end of this article.

We use the Dicke basis $\{|I,m,\beta\rangle\}$, where $I(I+1)$ is
the eigenvalue of the collective angular momentum operator
$\mathbf{I}^2$, the eigenvalue of $I^z$ is given by $m$, and $\beta$ is the
permutation quantum number \cite{ACGT72}.
The initial state of the spin bath in the following is the
identity
\begin{equation}
\rho_{\rm bath} = \frac{1}{2^N} \sum_{I,m,\beta} \proj{I,m,\beta}
= \one_{2^N}/2^N.\label{initialstate}
\end{equation}
In the following we omit $\beta$, which does not enter in the
dynamics.
%Its only effect is to give a weight $D(I)$ to the
%$(I,m)$-subspace.
This situation of a completely unknown bath state is, e.g., a
suitable description for GaAs QDs even at temperatures as low as
100 mK~\cite{PJT+05,KFE+05}. In the following, time will be given
in units of $N/A$. Even though the idea we present is applicable
to any (quasi-)static bath, we perform all estimations for GaAs,
i.e.~in particular $A^{-1}\approx 40$ ps.

\begin{figure}[ht]
\begin{center}
\hspace{0.3cm}
\includegraphics[width=0.34\columnwidth,height=4.4cm]{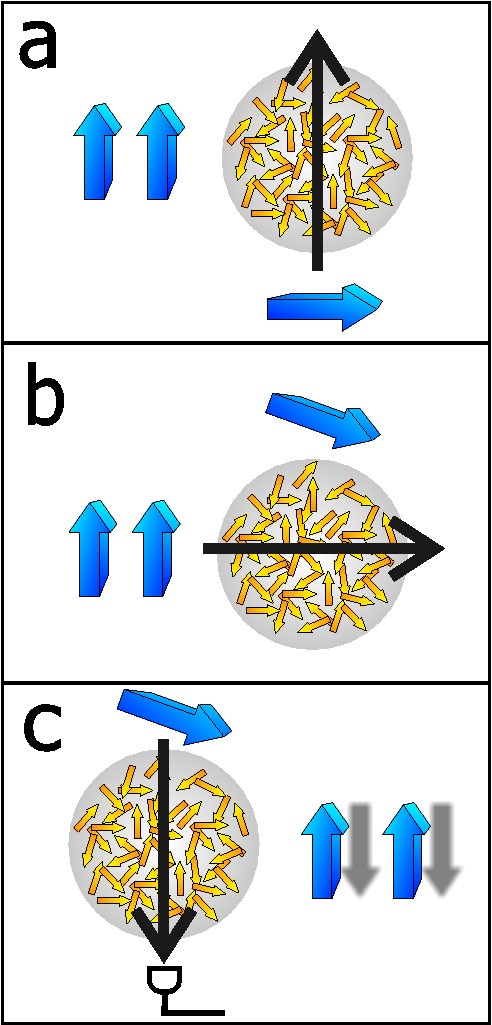}
\hspace{0.5cm}
\includegraphics[width=0.50\columnwidth,height=4.4cm]{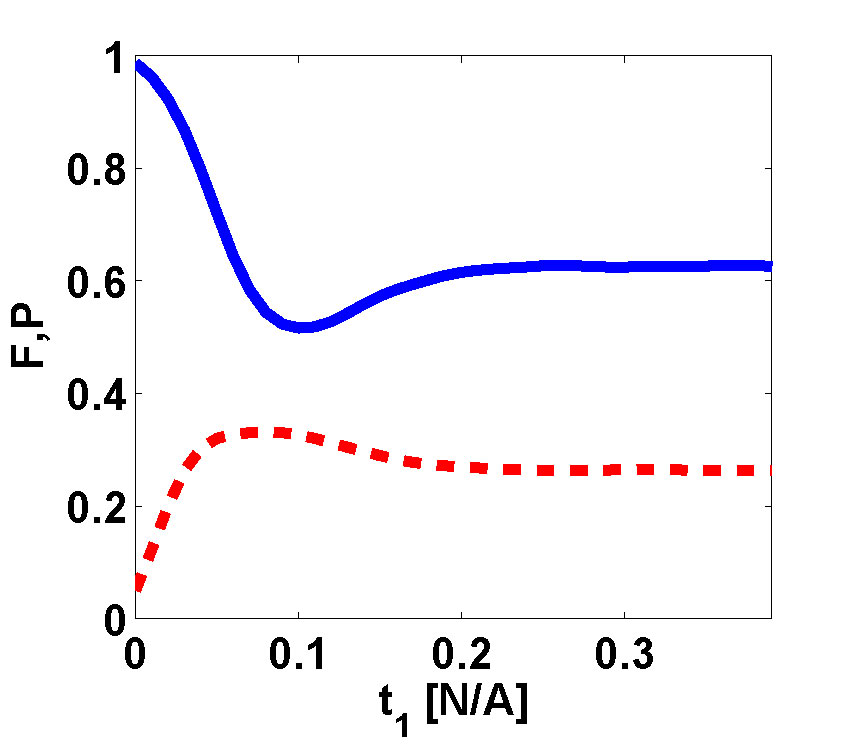}
\caption{(Color online) Left: Sketch of the protocol. (a) The $z$-polarized
  ``control electron'' interacts resonantly with the nuclear spin bath. (b) A
  sequence of $x$-polarized electrons interacts off-resonantly with the bath.
  (c) The control electron interacts resonantly again and is then measured in
  the $z$-basis.
  %Result $\downarrow$ projects the other electrons to a GHZ state.
  Right: Time dependence of overlap $F$ with Bell state $|\phi_-\rangle=(|++\rangle - |--\rangle)/\sqrt{2}$
  (solid blue line) for $N=10^3$. Dashed, red line shows the probability
  $P$ for a $\downarrow$ measurement.} \label{Fig1}
\end{center}
\vspace{-0.5cm}
\end{figure}
%---------------

The first electron spin (which we also refer to as ancilla
electron) is prepared in the state $|\up\rangle$ and interacts
\emph{resonantly} for a time $t_1$ with the nuclear spin bath
\begin{equation}
U | I,m,\uparrow \rangle = c_{Im}(t_1) | I,m,\uparrow \rangle +
s_{Im}(t_1) | I,m+1,\downarrow \rangle, \label{bla}
\end{equation}
with $U=e^{- i H t_1}$ and
\begin{align}
c_{Im}(t_1) &= \cos\left(\frac{(1+2I)t_1}{4}\right) - i
\frac{1+2m}{1+2I}
\sin\left(\frac{(1+2I)t_1}{4}\right), \nonumber \\
s_{Im}(t_1) &=
\frac{-2i\sqrt{(I-m)(1+I+m)}}{1+2I}\sin\left(\frac{(1+2I)t_1}{4}\right).\nonumber
\end{align}
Then the next electron spin, initial state $|+\rangle =
1/\sqrt{2}({|\up\rangle} + |\dn\rangle)$, interacts for a time $t_2$
\emph{off-resonantly} (e.g., in the presence of a large $B_z$) with the spin
bath. For $g^\ast \mu_B B_z\gg A/\sqrt{N}$, the flip-flop part of the
Hamiltonian can be approximately neglected~\cite{CoLo04}, yielding
\begin{displaymath}
V(t_2)|m,+\rangle = \frac{1}{\sqrt{2}}|m\rangle\big(e^{-i (\tilde B +
m) t_2/2}|\up\rangle + e^{+i (\tilde B + m) t_2/2}|\dn\rangle
\big),
\end{displaymath}
where $\tilde B = g^\ast \mu_B B_z N/A$ and the index $I$ has been
omitted for brevity. Remarkably, by choosing the interaction time
$t_2 = \pi$, the state of the electron is transformed to
$(-i)^m\ket{(-)^m}$, i.e. for even $m=2k$ to $(-1)^k|+\rangle$ and
for odd $m=2k+1$ to $-i(-1)^k|-\rangle$. For convenience we assume
$\tilde B t_2/2 = 2 \pi \ell$, $\ell\in \mathbb{N}$, which is
adjusted by the ``free'' parameter of the large field.

With the third electron, also in $|+\rangle$ initially and the
same interaction, the state becomes
\begin{equation}
\pm c_{Im}(t_1) | I,m,\uparrow \rangle |\pm\pm\rangle  \mp  s_{Im}(t_1) |
I,m+1,\downarrow \rangle|\mp\mp\rangle,\label{intermediatestate}
\end{equation}
with upper (lower) signs referring to even (odd) $m$.

In the final step, the ancilla electron interacts resonantly with
the nuclei again [cf.~Eq.~(\ref{bla})], giving
\begin{align}
\pm c_{Im}(t_1) \big|\pm\pm\big\rangle & \big[c_{Im}(t_1)| m ,
\uparrow \rangle + s_{Im}(t_1)|\hspace{-0.006cm} m
\hspace{-0.06cm} +\hspace{-0.06cm} 1,
\downarrow \rangle\big] \label{noeq}\\
  \mp  s_{Im}(t_1) \big|\mp\mp\big\rangle & \big[c_{Im}^\ast(t_1)
|m+1,\downarrow \rangle + s_{Im}(t_1)| m , \uparrow
\rangle\big],\nonumber
\end{align}
for even/odd $m$ and is eventually measured projectively in the
$z$-basis. If the measurement outcome is $\downarrow$ it is clear
from Eq.~(\ref{noeq}) that in each subspace the second and third
electrons are in the maximally entangled state
\begin{equation}\label{condstatehom}
| m + 1 \rangle \big( | + + \rangle - e^{\pm i \phi_m} | - -
\rangle \big)/\sqrt{2}, %\label{finalstate}
\end{equation}
where the phase $\phi_m=2\arg(c_{Im})$ depends on the quantum
numbers $I$ and $m$, leading to a washing out of the entanglement
when the average over the different subspaces is taken. However,
for short times ($(2I+1)t_1\ll1$ for typical values of
$I\sim\sqrt{N}$) this phase tends to zero and near ideal
entanglement is created, albeit at the price of a lower success
probability, see Fig.~\ref{Fig1}.

%---------------
\section{Multipartite Entanglement}
%-----------
The presented scheme generalizes in a straightforward manner to
multipartite entanglement creation. Following the same protocol
using $n$ electrons with arbitrary initial states
$|\psi_1\rangle,\ldots,|\psi_n\rangle$, the final state becomes
\begin{equation}
|\Psi_n\rangle = 1/\sqrt{2}\left( \one + [(-1)^{m+1} i\sigma_z]^{\otimes
n}\right)|\psi_1,\ldots,\psi_n\rangle,
\end{equation}
where the matrices are given in the standard $z$-basis and we
assumed the short time limit $t_1 \rightarrow 0$ for clarity. If
$|\psi_k\rangle = |+\rangle$ for all $k$, this is a $n$-partite
Greenberger-Horne-Zeilinger (GHZ) state. The $m$-dependent relative phase in
the above equation restricts to 
generation of GHZ-states with even particle number.

When multiple resonant interactions with the ancilla and varying
interaction times are allowed, a larger class of states becomes
accessible.  To see which states can the in principle be prepared,
we exploit the similarity of our setup to the sequential
entanglement generation scheme analyzed in Ref.~\cite{SSV+05}.
There it was shown, that all the matrix product states (MPS) of bond
dimension $d$ can be prepared if a string of qubits interacts
sequentially with a $d$-dimensional ancilla system and
\emph{arbitrary
  unitaries} can be performed on ancilla and qubit in every step.

To apply this result to the present case, ancilla electron and
nuclear spin system together represent the control qubit: an
effective $d=2$ system with Hilbert space spanned for given
$(I,m)$ by $\{|I,m,\uparrow\rangle, |I,m+1,\downarrow\rangle\}$.
To see that arbitrary unitaries are possible, note that
$x$-rotations of the control qubit are caused by resonant
interaction, while a static $B_z$-field causes $z$-rotations. From
these, all single qubit gates on the control qubit can be
constructed.  The off-resonant interaction considered before
performs essentially a CNOT gate between the passing and the
control qubit. In the CNOT gate the ``control qubit'' is the
control and the passing electron the target, in the
$|\up,\downarrow\rangle$ and $|\pm\rangle$ basis, respectively.
Combined with single-qubit gates (on the passing electron), this
seems to be enough to allow for arbitrary transformations on the
coupled control-target system. However, the situation is more
complicated since the effective gate performed by the off-resonant
interaction \emph{differs} for even and odd parity of the control
qubit, namely $V(\pi) = e^{(-1)^m i\frac{\pi}{4}
  \sigma_z}\sigma_{x,1}^mCNOT_{1\to2}\sigma_{x,1}^m$, i.e., there is not only,
as seen before, a parity-dependent phase but whether logical-0 or
logical-1 controls the bit-flip in the passing qubit also depends on
the parity of $m$. One way to remove this $m$-dependence and enable
the generation of arbitrary states is to perform an ``$I^z$ parity
measurement'' by sending an electron $|+\rangle$ into the dot, and
then measure it in the $|\pm\rangle$ basis after off-resonant
interaction for a time $\pi$. Depending on the outcome, either the
odd or the even states are projected out. Remarkably, gaining this
single bit of information about the $2^N$-dimensional bath then
allows us to remove all $m$-dependence and perform clean CNOT-gates.
Hence the interactions outlined above are sufficient to prepare all
$d=2$-MPS with high fidelity and, if the passing electrons can be
brought into interaction with the ancilla again at any time,
\emph{arbitrary} two-qubit gates can be performed, which implies
that all matrix product states with two dimensional bonds can be
sequentially created~\cite{SSV+05}.

Direct resonant interactions lead to very low fidelity
$x$-rotations due to averaging over the different subspaces,
indicating that prior measurement \cite{KCL06,SBGI06,GTD+06} or
cooling \cite{CCG07} of the spin bath might be necessary. More
sophisticated control schemes, however, allow for near unit
fidelity single qubit rotations with no prior preparation: In
Ref.~\cite{GRCi03} it was proven that high fidelity arbitrary
single qubit gates can be effected by a Hamiltonian $H=\delta
\sigma_z + \Omega (\sigma_x \cos\phi + \sigma_y \sin\phi )$, when
only the parameter $\phi$ can be controlled precisely. For
$\delta$ and $\Omega$ it is sufficient to know that they are
non-zero for some value of a controllable external parameter and
zero for another. In our situation we have the three Hamiltonians
$ H_1 = \Delta \sigma_z= B \mu_I \sigma_z /2$ (nuclear Zeeman),
$H_2 = \frac{A}{2N}[ (m+1/2)\sigma_z + \xi_{I,m} \sigma_x]$
(resonant HF), and $H_3 = \frac{A}{2N}[\pm(\tilde B + m+1/2) +
\tilde B \mu_I/(\mu_B g^\ast) ] \sigma_z$ (off-resonant HF) at
hand. The Pauli matrices are acting on the control qubit, $\mu_I$
is the nuclear magnetic moment and $\xi_{I,m} =
\sqrt{I(I+1)-m(m+1)}$. The plus and minus signs for $H_3$ can be
effected through spin flips of the passing electron (recall that
$\sigma_x e^{iHt} \sigma_x = e^{i\sigma_x H\sigma_x t}$ and
$\sigma_x \sigma_z \sigma_x = -\sigma_z$). These Hamiltonians can
be switched on and off (adiabatically \footnote{We
  assume here perfect storage of the ancilla electron in a ``protected''
  region free of HF interactions. HF Hamiltonians can then be switched
  adiabatically by shifting the electron wave function slowly from the
  protected region to the spin bath.}) at will. Appropriate iterations of
evolutions can lead to effective Hamiltonians of weighted sums and
commutators of $H_{1,2,3}$. In particular, the subspace
independence of the parameter $\Delta\propto B$ allows for
generation of any weighted sum of $\sigma_x$ and $\sigma_y$ with
the weights being $(I,m)$-independent, thus making the results of
Ref.~\cite{GRCi03} applicable. We have thus shown that while naive
use of resonant interactions will lead to poor gate fidelities for
the control qubit, enhanced control schemes still allow for full
access to high fidelity rotations. Hence, in principle, universal
quantum computation on an electron-spin quantum register can be
performed, with all interactions mediated by the highly mixed spin
bath.

%---------------
\section{Experimental Feasibility}
%------------
We discuss now various couplings that have been neglected in the
idealized Hamiltonian \Eqref{Hamiltonian} but are present in the
QD setup. We are concerned here only with their effects on the
basic entanglement generation scheme. It is clear that the scheme
can only work as long as it is fast compared to the electron $T_2$
time, since the coherence of the ancilla electron must be
preserved. We see below that neither nuclear dynamics nor
inhomogeneity place more stringent conditions on our scheme.

\subsection{Inhomogeneity}
The HF Hamiltonian in QDs has a slightly different form from the one
in Eq.~(\ref{Hamiltonian}), because the collective bath operators have
a spatial dependence $A/N I^\mu\to A^\mu \equiv \sum_i \alpha_i
I_i^\mu$, $\mu=\pm,z$. The coupling constants $\alpha_i$ are $\propto
\mu_{I,i} |\psi_e(r_i)|^2$, with $|\psi_e(r_i)|^2$ being the
probability of finding the electron at location $r_i$ and
$A=\sum_j\alpha_j$ denotes now the effective (average) hyperfine
coupling strength. We focus our analysis on short resonant interaction
times $\Delta t_1 \ll \sqrt{N}/A$. The electronic state after the
above protocol conditioned on a $\downarrow$ measurement is
proportional to
\[ \sum_{\psi,\psi'}\langle \psi' | A^+ e^{i (\sigma_1^z + \sigma_2^z)A^z
t_2} + e^{i (\sigma_1^z + \sigma_2^z)A^z t_2} A^+ |
\psi\rangle|\hspace{-0.04cm}++\rangle\times {\rm H.c.}, \] where
the Pauli matrices act on the off-resonant electrons and
$\ket{\psi}=\ket{i_1\dots i_N}, \ket{\psi'}=\ket{i_1'\dots i_N'}$ label the
orthonormal basis of $I^z_j$ eigenstates. Evaluating
the matrix elements and introducing the normalization we get
\begin{equation}\label{condstate}
\rho(t_2) = \frac{1}{\mathcal{N}(t_2)}\sum_j \alpha_j^2
\hspace{-0.2cm}\sum_{\substack{i_1,\ldots,i_N=\pm 1/2\\i_j=-1/2}}
\hspace*{-5mm}(|+_0 +_0 \rangle + |-_j -_j \rangle)\times {\rm
H.c.},\end{equation} with the states $|+_0(t_2)\rangle = e^{i
\omega_0
t_2/2}|\up\rangle +  e^{-i \omega_0 t_2/2 }|\dn\rangle$ and
$|-_j(t_2)\rangle = e^{i \omega_j  t_2/2}|\up\rangle +  e^{-i
\omega_j  t_2 /2}|\dn\rangle$, both of which depend on the nuclear
spin configuration $\{i\}$ via the the frequencies $\omega_0=\omega_0
(\{ i \})= \sum_\ell \alpha_\ell i_\ell$ and $\omega_j = \omega_0
+ \alpha_j$ and the normalization $\mathcal{N}(t_2) = \sum_j
\alpha_j^2 [3 + \cos(\alpha_j t_2)]$. The time dependence of the
states has been omitted for brevity in the above formula.
Straightforwardly, one now determines the fidelity $F(t_2)=\langle \phi_- |
\rho(t_2)|\phi_-\rangle$ with the desired maximally entangled state
$\ket{\phi_-}\sim\ket{+_0+_0}+\ket{-_0-_0}$ as
\begin{equation}\label{eq:fid}
F(t_2) = 2 \sum \alpha_j^2 \Big/ \sum_j \alpha_j^2(3 +
\cos(\alpha_j t_2)).
\end{equation}
This expression readily gives the fidelity for arbitrary particle
numbers and arbitrary distributions of coupling constants. For
$N\gg 1$, the obtained value is independent of particle
number, and for the relevant situation of Gaussian coupling $F=
0.90, 0.83, 0.78$ for 1D, 2D and 3D, respectively. Including the
difference in magnetic moments for Ga and As [$^{75}$As:
$\mu_{I,{\rm As}}=1.44$, $^{69}$Ga: $\mu_{I,{\rm Ga},1}=2.02$
(60\%) and $^{71}$Ga: $\mu_{I,{\rm Ga},2}=2.56$ (40\%) \cite{SKL03}] these
values become $F= 0.83, 0.78, 0.74$, indicating that our scheme
is not compromised by realistic inhomogeneities.

For small inhomogeneity we find the optimal time $t_2^{\rm (opt)}$ by
setting the time-derivative of $F$ zero and expanding the equation in
terms of the deviations $\epsilon_j=\alpha_j-\alpha^*$, where
$\alpha^*=\pi/t_2^{\rm (opt)}$. Going to second order in the small
parameters $\epsilon_j/\alpha^*$, the ensuing quadratic equation
yields $\alpha^*= \eta_1[5-\sqrt{1+24(1-\eta_2/\eta_1^2)}]/4$, with
$\eta_x = \frac{1}{N}\sum_j \alpha_j^x$.
%($=A_eff^x$ in the homogeneous limit).
% note used gamma to avoid confusion with cooling paper and yet get N out 
% of formula for \alpha* 
%Note that $\alpha^*$ is not equal to (but in fact slightly larger
%than) the average coupling strength $\eta_1$, effectively giving
%more weight to the more strongly coupled nuclei than the naive approach of coupling time $\pi/\eta_1$ would. 
Plugging $t_2^{\rm (opt)}$ back into Eq.~(\ref{eq:fid}) and keeping
terms up to second order we find $F(\pi/\alpha^*) =
1-\frac{\pi}{2N}\sum_j(\epsilon_j/\alpha^*)^2$.

\subsection{Nuclear Zeeman Energies}
For the times considered, nuclear Zeeman energies lead to an important
relative phase $B_z \mu_{I,j}t_2$ for each two terms in the sum of the
conditional state given in Eq.~(\ref{condstate}).

Considering one homogeneously coupled species of nuclear spins, the state of
Eq. (\ref{condstatehom}) will have an additional $m$-dependent phase. In each
invariant subspace this produces an overall phase $\propto B_z \mu_I t_2 m$,
and a relative phase $\propto B_z \mu_I t_2$ between the two parts of the
superposition. This might not seem harmful, but due to the parity effect, the
\emph{sign} of the phase depends on the parity of $m$. Since this phase is of
order $\pi$ it could spoil the protocol. However, by simply waiting for an
appropriate time $t_p$ after each of the $n$ electrons has passed the total
relevant phase is $(-−1)^m nB_z \mu_I (t_2 + t_p )$ and with $t_p + t_2$ an
integer multiple of $\pi/(\mu_I B_z)$ it is again $m$-independent. By the same
procedure, the nuclear Zeeman related phase can be removed for single-species
inhomogeneous systems.

For systems with strongly varying nuclear magnetic moments
$\mu_{I,j}$, the relative phase depends on ``which nuclear spin has
flipped'' and the waiting time no needs to be chosen such that all the
relative phases are close to $2k\pi$, otherwise the the final fidelity
may be strongly degraded. For the three species in GaAs this is the
case, e.g., for $B_z(t_2+t_p)\approx7\pi$, and assuming a flat wave
function still allows for a fidelity $\gtrsim0.9$ with moderate
overhead in time.  In principle, one can completely cancel the
undesired phase by removing the electrons from the QD and reversing
the magnetic field for $t_p=t_2$.

\subsection{Bath dynamics} The major internal dynamics of nuclear
spins in QDs stems from the indirect hyperfine mediated interaction,
and the direct dipolar interaction \cite{WiDa06,YLS05}.
Both mechanisms lead to bi-local errors that contain spin flip
terms $\propto (\Gamma_{d}^{kl} + \Gamma_{i}^{kl})I_k^+ I_l^-$ and
phase changing $zz$-terms $\propto\Gamma_{d}^{kl} I_k^z I_l^z$.
The transition rates for direct and indirect interactions are
$\Gamma_{dd}/r_{kl}^3$ and $\alpha_i \alpha_j/\Omega_e$,
respectively, where $\Omega_e$ is the electron Zeeman energy and
$r_{kl}= | \mathbf{r}_k - \mathbf{r}_l|$.

The dephasing interactions $\propto I^z_i I^z_j$ lead to a relative
phase between the terms in Eq.~(\ref{condstate}), similar to the
nuclear Zeeman energies. The energy difference, i.e. (in a mean
field treatment) the Zeeman splitting of a single nuclear spin in
the field of its neighbors, is a few times $\Gamma_{dd}$
\cite{CCG07}. Thus we need $N \Gamma_{dd}/(r_0^3 A)\ll 1$; given
$\Gamma_{dd}/r_0^3 \approx 0.1$ms \cite{SKL03} for nearest
neighbors, and $A\approx 40$ ps, this condition is readily fulfilled
even for large dots.

We have seen above that for each term in the mixed state, the qubits
rotate in the equatorial plane of the Bloch sphere with frequencies
$\alpha_j$ when the $j$th nuclear spin has been flipped. If this
particular spin is involved in a spin flip due to bath dynamics, the
resulting rotation with ``wrong'' frequency spoils the entanglement.
The errors in the rotation angle for the term containing the flip of
the $j$th spin are $\epsilon_{d,i}^j = \sqrt{\sum_k
(\Gamma_{d,i}^{kj} t)^2 (\alpha_j - \alpha_k)^2 (t_2^{\rm opt})^2}$,
and the final overall errors $\sum_j \alpha_j^2 \epsilon_{d,i}^j /
\sum_j \alpha_j^2$. We evaluate the above sums in the continuum
limit for Gaussian couplings and get for the indirect flips a total
error $\delta_i \pi^2 A \gamma_3^2 /(\gamma_4^2\Omega)$, where
$\delta_i$ is determined by the integrals over the coupling
constants. Taking $A/\Omega< 1$ for the large ($>1$ T) fields that
we require, we find errors $2.4\%$, $2.0\%$, and $ 1.5\%$ for 1D,
2D, and 3D, respectively, for $N=10^4$ (we define $N$ here as the
number of nuclei within the $1/(2e)$-width of the Gaussian). For the
direct nuclear dipole-dipole transitions the error is of size
$\delta_d \pi^2 \gamma_3^2\Gamma_{dd} /(A r_0^3 \gamma_4^2)$, where
numerical evaluation of the ``dipolar integrals`` $\delta_d$ yields
$0.01\%, 0.8\%, 5 \%$ for the same situation as above. This overall
error is thus on the order of a few percent for realistic
situations.

\subsection{Storage} We implicitly assumed the possibility of
storing the electrons protected from any bath. In QD structures this
could be achieved by shuttling the electrons to a nuclear spin-free
region or employing dynamical decoupling schemes, see
for example~\cite{WiDa06}. The required storage times of a few tens of
$\mu$s should be readily achieved.

\subsection{Imperfect Electronic Operations} A finite probability
that an up-electron is wrongly detected as a down electron (or vice
versa), degrades the final entanglement. However, as only one
electron needs to be measured, the effect is no worse for the
$n$-partite GHZ-state than for the Bell state $\ket{\Phi^+}$. The
same goes for variations in the resonant interaction time $t_1$.
%For small variations $\delta t_1$ in the short time limit the loss
%of entanglement scales quadratically in $\delta t_1/t_1$, and thus
%control on an order of magnitude less than $\sqrt{N}/A \sim 1-10$
%ns practically eliminates this error~. In
In contrast, errors in the electron preparation and variations in
the off-resonant interaction time $t_2$, since they affect each of
the $n$ electrons lead to a fidelity reduction that scales
exponentially with $n$. Variations in $t_2$ must be such that
$\tilde{B}\delta t_2\ll 1$ with $\tilde{B}\gg A/\sqrt{N}$, which
makes this the most stringent, but still
realistic~\cite{FHPD06,TED+05}, requirement for electron timing.

%---------------
\section{Summary and Conclusions}
We have considered the Heisenberg interaction of electron spin
qubits with a long-lived spin bath in a situation where \emph{nothing} is
known about the state of the bath and shown that nevertheless high fidelity
multipartite entanglement can be created via this bath.

The qubits do neither interact directly with each other nor
simultaneously with the bath at any time. Our protocol thus
demonstrates that even the interaction even with infinite temperature
systems can mediate highly coherent operations and thus represent a
valuable resource for quantum information processing that merits
further inverstigation.

In fact, when only one bit of information is extracted from the spin
bath, arbitrary gates between the bath and the qubits are possible,
and all matrix product states with 2D bonds can be created by
sequential interaction.

The explicit protocols we presented can be realized in quantum dot
setups and would (in typical GaAs dots) allow for the creation of
entanglement between two electrons on a timescale of a few $\mu$s.

\begin{acknowledgments}
  This work was supported by the DFG within SFB 631 and the
excellence cluster NIM.
\end{acknowledgments}

%----------------------------------------
%----------------------------------------
%\bibliographystyle{apsrev_short}
%\bibliography{AllBib}

\end{document}